# AUTOMATED OPERATION OF THE APS LINAC USING THE PROCEDURE EXECUTION MANAGER

R. Soliday, S. Pasky, M. Borland, Argonne National Laboratory, Argonne, IL 60439, USA

## Abstract

The Advanced Photon Source (APS) linear accelerator has two thermionic cathode rf guns and one photocathode rf gun. The thermionic guns are used primarily for APS operations while the photocathode gun is used as a free-electron laser (FEL) driver. With each gun requiring a different lattice and timing configuration, the need to change quickly between guns puts great demands on the accelerator operators. Using the Procedure Execution Manager (PEM), a software environment for managing automated procedures, we have made start-up and switchover of the linac systems both easier and more reliable. The PEM is a graphical user interface written in Tcl/Tk that permits the user to invoke 'machine procedures' and control their execution. It allows construction of procedures in a hierarchical, parallel fashion, which makes for efficient execution and development. In this paper, we discuss the features and advantages of the PEM environment as well as the specifics of our procedures for the APS linac.

## 1 TCL/TK CODE

### 1.1 PEM

The Procedure Execution Manager (PEM) is a graphical user interface tool that allows the user to execute Tcl/Tk machine procedures and monitor their progress (see Figure 1) [1]. At the Advanced Photon Source (APS), PEM procedures are used routinely during the operations of the different accelerators including the linac. A key advantage of the PEM is that it can be easily expanded by adding new machine procedures without changing the familiar user interface. The PEM allows the user to select an execution mode: Automatic, Semi-Automatic, or Manual. These levels signify the amount of interaction and monitoring that will occur. The machine procedures include 'steps' at which the PEM can pause. Manual mode pauses at all steps. Semi-Automatic only pauses at the first step. Automatic runs without pausing at any of the steps. As shown in the figure, a collection of routines can be grouped together and given a title. The machine procedures shown here are specifically designed for linac operations. Additional PEM screens exist for all of the accelerators at the APS. Since it was first written in 1996, the PEM has proven to be a very reliable and useful program.

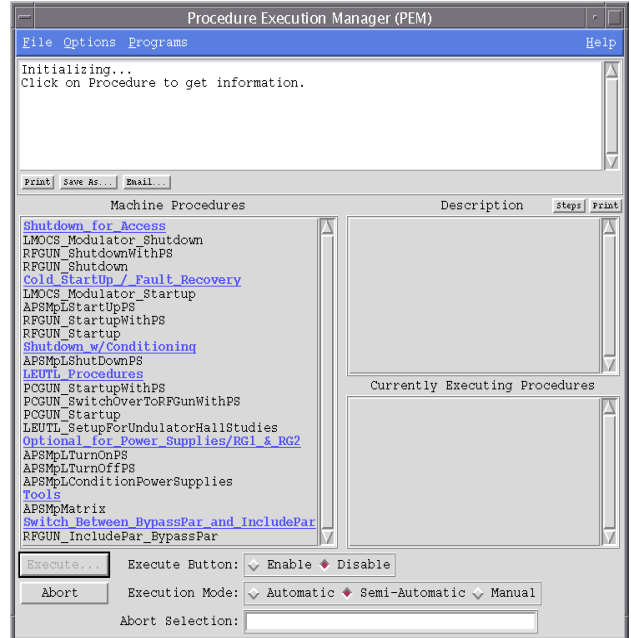

Figure 1: Procedure Execution Manager

### 1.2 Machine Procedures

A machine procedure is a Tcl/Tk procedure that follows a particular format where certain utility functions must be called from within the procedure [2]. These procedures are loaded into the PEM by using Tcl/Tk's built-in auto loading feature. The PEM then accesses a configuration file that lists the machine procedures that can be executed during a session. A simple example is shown here:

```
proc APSMpWriteThis {args} {
  APSMpStep "Writing to file"
  set fd [open /tmp/stuff w]
  puts $fd "$args"
  close $fd
  APSMpReturn ok "data written"
}
```

This procedure writes the value of 'args' to a file. Note that the last statement is APSMpReturn. This must be used in place of the return statement in all machine procedures.

For each machine procedure there can also be a companion description procedure defined as shown below. The return value from the description procedure is displayed in the 'Description' frame of the PEM tool when the corresponding procedure is selected.

```
proc APSMpWriteThisInfo {} {
  return "Place description of procedure here."
}
```

Machine procedures are designed to be executed primarily from within the PEM. The PEM takes full advantage of the format of a procedure to permit monitoring and controlled execution. These machine procedures may also be executed from any APS Tcl/Tk library, which allows them to be executed like any other Tcl/Tk procedure.

*1.3 Installation*

In order to run the PEM, Tcl/Tk with the Tcl-DP extension must be installed. The latest versions are recommended. The APS Tcl/Tk library and the PEM packages must also be installed. All of these are located on the OAG web site[1].

## 2  LINAC APPLICATIONS

The linear accelerator has three guns: two thermionic rf guns mainly used to support APS injections and one photocathode rf gun for experimental projects. All of these guns are important to APS as well as to future user demands. Normally experimental projects are parasitically operated during user beam mode. If stored beam is lost, it is important to the Operations Group to have a fast and reliable transfer from the experimental project to the APS injection configuration. This fast and reliable method of switching to and from the thermionic guns has been accomplished using the PEM tools.

In the past, operators manually initiated and monitored all systems involved in the switchover using Motif Editor and Display Manager (MEDM) screens. Since the linac has a multitude of MEDM screens that control every aspect of operations, the switchover was not an easy task. Operators normally had to switch back and forth between many MEDM screens as they worked. As demands on the operators increased due to system changes, some small shell scripts where written to perform tasks automatically. Although the scripts worked well, they were not always reliable because changes to machinery and operational procedures were being made without warning. Also most of these scripts were not regulated and did not have much of an error checking ability. When configured properly, PEM procedures follow the same steps an operator would take during equipment start-up. The PEM tools not only repeat steps faster, they also provide reproducibility.

Using PEM tools for linac operations, the operator no longer has to open multiple windows and work on one task at a time. Instead, the PEM is able to efficiently use multitasking to alleviate the burden on the operators in what can often be a stressful situation. By making the interface of the PEM simple and consistent, new machine procedures can be added. Operators can read the corresponding description and view the steps of a procedure to become familiar with it. This is not intended to reduce operator training, however; it merely acts as an additional source of information that may be valuable to operators.

## 3  PEM DEVELOPMENT

There are several different types of machine procedures ranging from simple low-level procedures to a collection of many submachine procedures. Often a machine procedure is a collection of smaller core procedures linked together either in series or in parallel. In doing so, we were allowed to work in a simple progression. Core procedures are the lowest level procedures that do the actual work and are written first. Once these core procedures have been tested, larger parent procedures are formed that call on all of the necessary machine procedures. This process is repeated many times, creating many levels of parent procedures. Table 1 shows an example of the machine procedures called when the photocathode start-up procedure is used. There are more layers than this table shows, but they are too numerous to display.

Table 1: Machine procedures called from the photocathode start-up procedure

| |
|---|
| 1. Start up photocathode gun with power supplies<br>    a.  Start-up power supplies<br>          i.  Turn off unused power supplies<br>          ii.  Turn on needed power supplies<br>          iii.  Condition power supplies<br>          iv.  Wait for conditioning to finish<br>    b.  Start up photocathode gun<br>          i.  Shut down thermionic gun<br>          ii.  Bring modulators to standby<br>          iii.  Bring modulators back up |

Using a modular method, the PEM can decrease the execution time by taking two or more nonsequential procedures and running them in parallel. These parallel procedures can split off indefinitely into subparallel procedures. All procedural steps report back to the original PEM and communicate the steps as they occur. This causes an interleaving of the steps displayed by the PEM as the execution progresses. Eventually, prior to exiting or executing additional steps, the parallel procedures must be joined. This joining ensures that all steps have been completed successfully prior to continuing with the program.

Another advantage of a modular method is the ability to make a change in a core procedure. If this machine procedure is called by multiple machine procedures, the change will affect all of them. This means that multiple source codes do not need to be changed if a design or operation change is implemented that requires a change in a core machine procedure.

---

[1] http://www.aps.anl.gov/asd/oag/oaghome.shtml

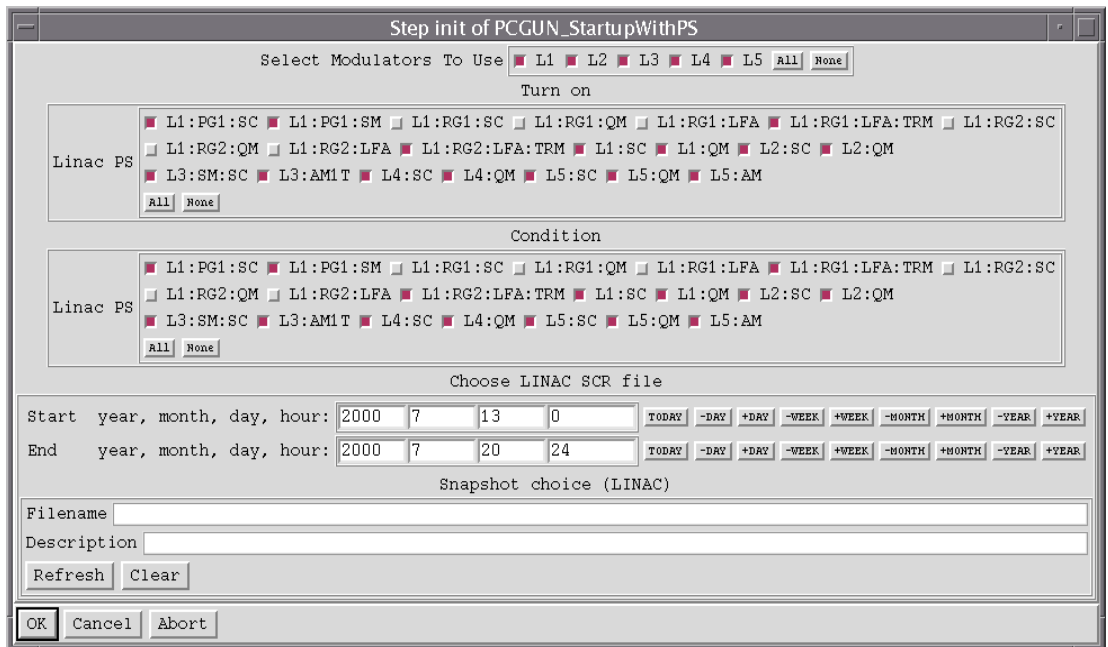

Figure 2: Dialog for photocathode gun start up with power supplies

The dialog screen shown for the switchover to the photocathode gun (see Figure 2) allows the operator to select a snapshot file to be restored at the end of a switchover. A snapshot file is a database file that includes all of the lattice power supply settings needed to reproduce the same beam as when the snapshot was recorded. Once executed, the PEM procedure opens another display window that shows each step as it occurs. The particular procedure shown in Table 1 will start up and condition the power supplies and in parallel start up the photocathode gun. The subprocedures for the power supply start up and photocathode gun start up are run in series. Because this machine procedure is able to do many tasks at once as well as perform many safety checks, it allows the operator to attend to many other tasks.

When an unexpected condition occurs, the PEM displays a dialog box to the operator containing a description of the problem and requests that the operator attempt to fix it manually. Buttons for continuing and/or aborting are often displayed on these dialog boxes. An abort button is always displayed on the PEM screen during execution for those situations when something may go wrong and continued operation of the PEM may be unnecessary or unwise. Along these same lines, a log daemon is used with the PEM to log any and all error messages that may occur during normal operations. This has been used to track down some obscure problems that occur infrequently.

## 4  CONCLUSION

Without the use of the PEM, the multiple operating modes of the linac at the APS would not be possible. The PEM has proven itself to be a useful tool capable of handling a wide array of tasks. Switching between the operating modes with the assistance of the PEM has been made almost trivial for the operator. Without the assistance of the PEM, the method for switching operating modes would require detailed knowledge of all the systems, execution of all the switchover steps in the correct order, and a lot of time.

## 5  ACKNOWLEDGMENT

PEM was implemented by C.W. Sanders, formally of APS, based on concepts developed by M. Borland and C.W. Sanders. This work is supported by the U.S. Department of Energy, Office of Basic Energy Sciences, under Contract No. W-31-109-ENG-38.